\documentclass[sigconf, screen]{acmart}

\usepackage{amsmath}
\usepackage{multirow}
\usepackage{pbox}
\usepackage{caption}
\usepackage{subcaption}
\usepackage{atbegshi,picture}

\definecolor{DarkBlue}{HTML}{00008B}
\definecolor{green}{rgb}{0.0, 0.65, 0.31}
\definecolor{purple}{HTML}{ae017e}
\definecolor{mediumblue}{rgb}{0.0, 0.0, 0.8}

\newif{\ifhidecomments}
  \hidecommentsfalse 
\ifhidecomments
    \newcommand{\chelsea}[1]{}
    \newcommand{\xuhui}[1]{}
    \newcommand{\hong}[1]{}
    \newcommand{\maarten}[1]{}
    \newcommand{\jodi}[1]{}

\else
    \newcommand{\chelsea}[1]{\textbf{\small\sffamily{\textcolor{purple}{[#1 -- Chelsea]}}}}
    \newcommand{\hong}[1]{\textbf{\small\sffamily{\textcolor{DarkBlue}{[#1 -- Hong]}}}}
    \newcommand{\xuhui}[1]{\textbf{\small\sffamily{\textcolor{green}{[#1 -- Xuhui]}}}}
    \newcommand{\maarten}[1]{\textbf{\small\sffamily{\textcolor{red}{[#1 -- Maarten]}}}}
    \newcommand{\jodi}[1]{\textbf{\small\sffamily{\textcolor{orange}{[#1 -- Jodi]}}}}
  \fi

\AtBeginDocument{%
  }

\setcopyright{cc}
\copyrightyear{2025}
\acmYear{2025}
\acmDOI{XXXXXXX.XXXXXXX}
\acmConference[HEAL @ CHI'25 Workshop]{Human-centered Evaluation and
Auditing of Language Models Workshop at CHI 2025}{April 26th, 2025}{Yokohama, Japan}
\acmISBN{978-1-4503-XXXX-X/2025/04}

\begin{document}

\title{Rethinking Theory of Mind Benchmarks for LLMs: \\ Towards A User-Centered Perspective}
\renewcommand{\shorttitle}{Rethinking Theory of Mind Benchmarks}


\author{Qiaosi Wang}
\email{qiaosiw@andrew.cmu.edu}
\affiliation{%
  \institution{Carnegie Mellon University}
  \city{Pittsburgh}
  \state{PA}
  \country{USA}
}

\author{Xuhui Zhou}
\email{xuhuiz@cs.cmu.edu}
\affiliation{%
 \institution{Carnegie Mellon University}
 \city{Pittsburgh}
  \state{PA}
  \country{USA}
}

\author{Maarten Sap}
\email{maartensap@cmu.edu}
\affiliation{%
 \institution{Carnegie Mellon University}
 \city{Pittsburgh}
  \state{PA}
  \country{USA}
}

\author{Jodi Forlizzi}
\email{forlizzi@cs.cmu.edu}
\affiliation{%
 \institution{Carnegie Mellon University}
 \city{Pittsburgh}
  \state{PA}
  \country{USA}
}

\author{Hong Shen}
\email{hongs@andrew.cmu.edu}
\affiliation{%
 \institution{Carnegie Mellon University}
 \city{Pittsburgh}
  \state{PA}
  \country{USA}
}

\renewcommand{\shortauthors}{Wang et al.}

\begin{abstract}
The last couple of years have witnessed emerging research that appropriates Theory-of-Mind (ToM) tasks designed for humans to benchmark LLM’s ToM capabilities as an indication of LLM’s social intelligence. However, this approach has a number of limitations. Drawing on existing psychology and AI literature, we summarize the theoretical, methodological, and evaluation limitations by pointing out that certain issues are inherently present in the original ToM tasks used to evaluate human's ToM, which continues to persist and exacerbated when appropriated to benchmark LLM's ToM. Taking a human-computer interaction (HCI) perspective, these limitations prompt us to rethink the definition and criteria of ToM in ToM benchmarks in a more dynamic, interactional approach that accounts for user preferences, needs, and experiences with LLMs in such evaluations. We conclude by outlining potential opportunities and challenges towards this direction.
\end{abstract}




\maketitle

\section{Introduction}
\begin{figure}
    \centering
    \includegraphics[width=0.8\linewidth]{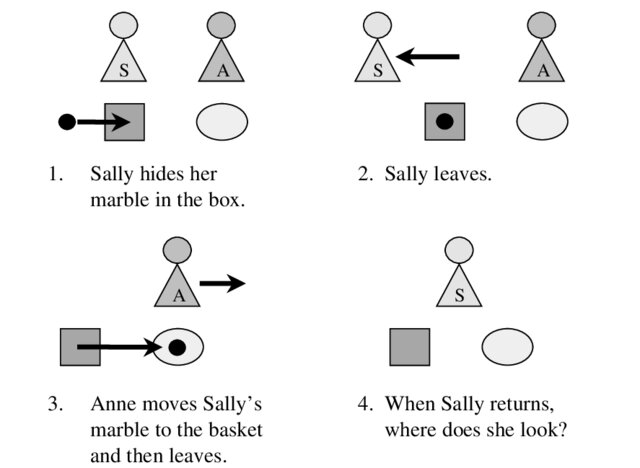}
    \caption{An illustration of the Sally-Anne test commonly used to evaluate children's Theory of Mind. Figure reproduced from~\citeauthor{scassellati2001foundations} [2001].}
    \label{fig:sally-anne}
\end{figure}

In recent years, Theory of Mind (ToM) has gained much attention in the evaluation and benchmarks of Large Language Models (LLMs) due to its fundamental role in social cognition. ToM is the human social and cognitive capability of attributing mental states (e.g., knowledge, intentions, desire, emotions) to ourselves and others based on observable behavioral and verbal cues, with the goal of predicting and making sense of others' actions~\cite{baron1985does,baron1999evolution,premack1978does}. Many human social behaviors are enabled by ToM, such as persuasion, teaching, repairing communication breakdowns, building shared plans and goals~\cite{baron1999evolution}, all of which requires us to make conjectures about what's going on in others' minds (e.g., their intentions, knowledge, preference, motivations) to behave accordingly and achieve optimal social interaction outcomes. Given its fundamental role in human social interaction, ToM has been studied extensively across various disciplines especially in developmental and clinical psychology, where researchers have studied the emergence and development of ToM in children as well as the role of ToM in people with autism or schizophrenia who tend to experience difficulty in social interactions with neurotypical people\footnote{see \href{https://en.wikipedia.org/wiki/Double_empathy_problem}{the Double Empathy problem}}~\cite{milton2012ontological,wellman2018theory,rakoczy2022foundations,baron2000theory}. Throughout these research endeavors, researchers have come up with a number of tasks to assess people's ToM capability. One of the most famous ToM tasks is perhaps the Sally-Anne test (as illustrated in Fig.~\ref{fig:sally-anne}), which presents a scenario to assess the children's understanding of false beliefs, an important indication of the child's ToM ability in recognizing others can have beliefs that the child know to be false. 

Recently, such ToM tasks have been appropriated as benchmarks to evaluate LLM's ToM capability, with the goal of assessing their social cognitive abilities. Some studies have directly applied these human-intended tasks to LLMs and drawn bold claims based on model performance on these ToM tasks~\cite{kosinski2023theory,bubeck2023sparks}. For example,~\citeauthor{kosinski2023theory} [2023] claimed that ``ToM may have spontaneously emerged in LLMs'' after models passed over 70\% of false belief tasks. Similarly,~\citeauthor{bubeck2023sparks} [2023] concluded that GPT-4 demonstrates ``a very advanced level of ToM'' based on its superior performance in false belief, emotion recognition, and intention inference tasks compared to other models.  
These claims have sparked lively debate within the AI community. In response, there is growing recognition that evaluating ToM in LLMs requires benchmarks grounded in NLP evaluation methods and tailored to the unique affordances and limitations of LLMs, rather than repurposing cognitive assessments designed for humans. Following this notion, many work has created variations based on the existing ToM tasks' structure and content to benchmark LLM's ToM capability~\cite[e.g.,][]{xu2024opentom,kim2023fantom}. Others have pointed out the concerning robustness of drawing claims based on LLM passing human-intended ToM tasks by demonstrating LLM failures when these tasks were modified with trivial alterations~\cite[e.g.,][]{ullman2023large,shapira2023clever,sap2022neural}. 





Responding to the growing call for examining the limitations of appropriating human-intended ToM tasks to benchmark LLM's ToM~\cite{shapira2023clever,sap2022neural,ullman2023large}, this position paper aims to surface the number of limitations embedded in the original human-intended ToM tasks--- limitations that not only persisted but are amplified when these tasks are repurposed to evaluate LLMs. These inherited limitations cast doubt on the validity of conclusions drawn about LLM's ToM and social capability based on this type of evaluations~\cite{shapira2023clever}. In this paper, we first provide an overview of the various ToM tasks used to evaluate humans and LLMs based on existing work. Drawing from both psychology and AI literature, we summarize the theoretical, methodological, and evaluation limitations of appropriating ToM tasks as benchmarks for LLM's ToM capabilities. Building on this foundation, we take an HCI perspective to rethink why, what, and how we benchmark LLMs’ ToM capabilities, highlighting potential opportunities and challenges for designing user-centered ToM evaluations.


\section{Evaluating Theory of Mind in Humans and Large Language Models}

\subsection{Assessing Human ToM through ToM Tasks}
While~\citeauthor{premack1978does}[1978] did not specify what ToM encompasses when they coined the term, decades of psychology literature has established ToM as a multi-faceted construct that includes various cognitive and affective dimensions.~\citeauthor{fu2023systematic}[2023] distillled four construct dimensions of ToM from a systematic literature review of 127 ToM measures: cognitive-interpersonal, cognitive-intrapersonal, affective-interpersonal, and affective-intrapersonal. Similarly,~\citeauthor{beaudoin2020systematic} [2020] identified 220 ToM measures used to evaluate children's ToM and pinpointed seven dimensions of ToM: emotions, desires, intentions, percepts, knowledge, beliefs, and mentalistic understanding of non-literal communication. These dimensions are further divided into 39 types of ToM sub-abilities. However, given the broad definition of ToM as the ability to attribute mental states to self and others, researchers have pointed out the nonspecificity of ToM, which can be used to simultaneously include different cognitive constructs such as emotional reactivity and facial expression categorization~\cite{quesque2020theory}. This has led to construct validity issues of certain ToM tasks not actually measuring ToM (e.g., Reading the Mind in the Eyes Test). 

Over the years, hundreds of ToM tasks have been proposed and used to evaluate human's ToM capability. In developmental psychology, ToM tasks have been used to identify developmental milestones as well as understanding social deficits in children with autism spectrum disorders~\cite{beaudoin2020systematic,baron1985does}. While ToM tasks can be administered in varying presentation modes, a typical ToM task often comprises of a social scenario in the form of a story, a comic, or even a video, followed by questions (typically multiple-choice) to the child about the mental state of the characters in the social scenario~\cite{beaudoin2020systematic,fu2023systematic}. Besides the classic Sally-Anne false belief task (shown in Fig.~\ref{fig:sally-anne}), other classic ToM tasks include faux pas, strange stories, second-order false belief, and more~\cite{hayward2017reliability}. Specifically, the faux pas task presents the child with a story, where one character makes a social mistake (e.g., saying that the dish cooked by the dinner host is not good in front of the host), and the child was asked if the character's behavior is appropriate and why~\cite{baron1999recognition}. The strange stories task presents the child with short stories of characters acting strangely by pretending, joking, or lying, and the child must infer the mental state of the characters to explain their behavior~\cite{happe1993communicative}. These tasks have far-reaching influences as many have been adapted and extended to measure ToM capabilities beyond children, and more recently, measuring LLMs' ToM.

\subsection{Current State of ToM Benchmarks for LLMs}
Much like how ToM tasks are administered to children, most ToM benchmarks for LLMs follow a similar approach: presenting static, text-based social scenarios and prompting the models to infer the reality and mental states of the characters~\citep{sap2022neural}. Model performance is typically scored based on answer accuracy~\citep{shapira2023clever}. These synthetic scenarios are largely appropriated from ToM tasks originally designed for humans and are adopted by several prominent and highly-cited ToM benchmarks. For example, inspired by the Sally-Anne test,~\citeauthor{le2019revisiting}[2019]'s dataset contains over 1000 distinct stories and questions prompting for the character's memory, reality, and false-beliefs. Similarly,~\citeauthor{kosinski2023theory} [2023]'s ToM benchmark contain 40 tasks focusing exclusively on false-belief scenarios.~\citeauthor{shapira2023well} [2023] used human experts and ChatGPT to generate synthetic social scenario stories based on the faux pas test.~\citeauthor{chen2024tombench} [2024] constructed eight scenarios covering a range of ToM dimensions to build a more comprehensive benchmark. These scenarios are often paired with multiple-choice questions, where only one option is deemed correct regarding the character’s mental state or relevant situational details. Such benchmarks are frequently reused or adapted in subsequent studies to evaluate LLMs’ ToM capabilities~\cite[e.g.,][]{shapira2023clever,sap2022neural}, and have had a lasting influence on how claims about LLMs’ social reasoning are generated.

Going beyond human ToM tasks, other ToM benchmarks also consist of similar format in presenting synthetic social scenarios to LLMs, followed by question-answering to gauge LLM's understanding of the social scenarios. For example,~\citeauthor{sap2019socialiqa}[2019]'s SocialIQA benchmark contains 38,000 multiple-choice questions about the intents and reactions to daily social interaction scenarios created through crowdsourcing. To better align with real-world scenarios, recent work has either adopted or generated natural real-world human-human conversation dataset to evaluate LLM's ToM capability through question-answering~\cite[e.g.,][]{soubki2024views,chan2024negotiationtom,kim2023fantom}--- some work has gone a step further to examine LLM's applications in using mental state inferences to predict and judge observable behaviors~\cite{gu2024simpletom}. While ToM benchmarks containing text-based static social scenarios provided vast convenience and accessibility for researchers to easily assess LLMs' ToM capability, other work has offered new opportunities to assess more dynamic and interactive social interactions in AI.~\citeauthor{zhou2023sotopia}[2023] presented SOTOPIA, an open-ended environment that can simulate and evaluate social interactions between artificial agents across a wide variety of social scenarios.~\citeauthor{chiang2024chatbot}[2024] created Chatbot Arena, an open platform that can involve humans to crowdsource questions to evaluate LLMs.~\citeauthor{shi2024muma}[2024] presented a multi-modal ToM benchmark to simulate the embodied, multimodal, multi-agent social interactions by providing video and text descriptions of people's behaviors in realistic household environment. 

The growing body of literature on empathy benchmarks for LLMs~\cite{sorin2024large,huang2025apathetic,welivita2024large,zhangdeveloping} offers valuable insights for ToM benchmark design, given that empathy is inherently a sub-construct of ToM—-- requiring the understanding and response to others' mental and emotional states. While most empathy benchmarks follow a similar structure to ToM benchmarks by presenting social scenarios to LLMs, many also involve human annotators to provide baseline comparisons for evaluating the empathy expressed in LLM-generated responses. For example,~\citeauthor{welivita2024large} [2024] engaged 1,000 participants to compare the empathetic quality of LLM-generated and human-generated responses.~\citeauthor{zhu2024reading} [2024] compared LLM's capability in generating empathic inferences about users' underlying goals and needs from product reviews against a baseline established by human designers. Although these benchmarks involve direct human participation, they typically position humans as a baseline for comparing LLM performance, leaving room to explore opportunities in incorporating user perspectives into the evaluation of LLM ToM capability.

\section{The Limitations of Appropriating ToM Tasks as LLM Benchmarks}
Although appropriating ToM tasks as LLM benchmarks has been a common practice in the AI benchmark literature, it was not until recently that this practice was put under scrutiny for producing rather controversial claims like ``ToM has spontaneously emerged in LLMs''~\cite[e.g.,][]{kosinski2023theory,bubeck2023sparks}. Outside of AI literature, however, decades of psychology research on the evaluation of human ToM has revealed certain limitations on ToM tasks, even when used to evaluate human ToM. Some of these limitations naturally persisted when ToM tasks are appropriated to measure LLMs' ToM. In this section, we summarize and consolidate the key limitations of appropriating ToM tasks as LLM benchmarks based on existing psychology~\cite[e.g.,][]{beaudoin2020systematic,quesque2020theory,ahmadi2015validity} and AI literature~\cite[e.g.,][]{ullman2023large, ma2023towards,shapira2023clever,sap2022neural} from three aspects: theoretical limitation, methodological limitation, and evaluation limitation. 

\subsection{Theoretical Limitation}
\textbf{ToM is a multi-faceted construct, yet it has been mostly measured on one dimension.} In the Psychology literature, several systematic literature reviews have pointed out the issue of ToM tasks only measuring one dimension, specifically ToM-beliefs via various false-belief tasks, with few providing comprehensive measures~\cite{beaudoin2020systematic,fu2023systematic,ahmadi2015validity}. In their review of ToM tasks used to evaluate young children,~\citeauthor{beaudoin2020systematic} [2020] pointed out that a whopping 75.5\% among the 220 ToM measures they identified focused solely on Beliefs (i.e., informational states that people believe to be true~\cite{ma2023towards}), whereas other ToM dimensions such as Emotions (i.e., emotional or affective states that people experience), Desires (i.e., human desires and wants without committed actions), Intentions (i.e., goals and intentions with committed actions) received far less attention (each accounted for 4.3\% to 23.9\% of the studies identified). Similarly,~\citeauthor{fu2023systematic}[2023] were only able to identify four out of the 127 ToM measures for children that cover all construct dimensions of ToM. 

Given the wide adoption of human-intended ToM tasks to evaluate LLM ToM, this phenomenon was also observed AI literature~\cite{ma2023towards,ying2025benchmarking}.~\citeauthor{ma2023towards} [2023] surveyed recent LLM ToM literature and observed ``an overwhelming research focus on intention and belief aspects of machine ToM'' yet other ToM aspects received little attention~\cite{ma2023towards}. For example, prominent ToM benchmarks such as~\citeauthor{le2019revisiting} [2019] were largely inspired by the Sally-Anne false belief tasks;~\citeauthor{kosinski2023theory} [2023]'s test set specifically included only variations of false belief tasks;~\citeauthor{shapira2023well} [2023]'s benchmark, while not focusing on ToM-Beliefs, was also inspired by one ToM task, the faux pas test, that measures only one dimension of ToM--- the ability to recognize social gaffe situations. While these benchmarks helped offer valuable insights into certain aspects of LLM's ToM capability, it is important to understand that these tasks only measure one dimension of the multi-faceted construct of ToM, and hence~\textit{provide limited insights and evidence in making claims about LLM's overall ToM capability}~\cite{ma2023towards,ying2025benchmarking}. 

\subsection{Methodological Limitation}
\textbf{Many ToM tasks lack construct validity and present mixed or lack of reports of test reliability.}~\citeauthor{premack1978does}[1978] first defined ToM as one's ability to attribute/impute mental states to self and others with the goal of predicting actions~\cite{premack1978does}, which has been widely agreed-upon and adopted by researchers across disciplines. However, this definition lacks specificity in the exact cognitive processes required to generate ToM-enabled behaviors. As a result,~\textit{many ToM tasks lack construct validity and can be solved through alternative low-level cognitive strategies} such as pattern recognition or learned association without requiring the participant to engage in mental state attribution when solving ToM tasks~\cite{quesque2020theory}. For example, one of the most used ToM tasks in examining human emotional ascriptions, the Reading the Mind in the Eyes test~\cite{baron2001reading}, measures emotion recognition rather than ToM~\cite{oakley2016theory,quesque2020theory}. While most of such invalid ToM tasks require input and output modalities beyond text (e.g., emotional attribution based on facial expressions) and haven't been used to benchmark LLM's ToM yet, AI researchers already uncovered LLMs leveraging similar tactics to pass ToM tasks---~\citeauthor{ullman2023large}[2023] found that LLMs fail on trivial alterations to ToM tasks,~\citeauthor{shapira2023clever}[2023] pointed out that LLMs rely on shortcuts, heuristics, and spurious correlations to pass ToM tasks,~\citeauthor{kim2023fantom}[2023] found that LLM ToM reasoning often appears illusory, as models can fail low-level reasoning questions despite correctly recognizing Beliefs. As multi-modal LLMs are being developed to process a variety of inputs beyond texts, researchers should take caution when using existing visual-based ToM tasks to benchmark LLM. In addition, as several AI researchers pointed out that given LLMs are typically trained on data that is readily available on the internet, ToM tasks that have been around for decades might have already been part of the models' training data, leading to~\textit{data contamination issues}~\cite{ma2023towards,ullman2023large} which makes it even more difficult to verify the validity of LLM's claimed ToM capability. 

In the past decade, hundreds of ToM tests have been created by psychologists~\textit{without reporting important psychometric properties to assess the validity and reliability of the measures (e.g., internal consistency, test-retest reliability)}~\cite{beaudoin2020systematic,fu2023systematic}.~\citeauthor{beaudoin2020systematic}[2020] found that only 20.2\% of their included ToM measures provided any such information;~\citeauthor{ahmadi2015validity}[2015] noted that only six of the 11 included ToM measures had been examined for construct validity;~\citeauthor{hayward2017reliability}[2017] identified notable validity and reliability issues of several prominent ToM measures (e.g., second-order false belief test and Strange Stories test has undesirable internal consistency). This has resulted in globally poor replicability of ToM measures in empirical studies~\cite{beaudoin2020systematic}. Similarly, AI researchers have used human annotators, generative AI, or both to create their own ToM tests. While this enables rapid and large-scale test generation to stress-test LLMs, the process is often opaque and inconsistently reported in terms of validity and reliability. Although inter-rater reliability among the human annotators is commonly noted, the specific processes and measures researchers took to ensure human annotators' correct understanding of the ToM dimensions, internal consistency of the hundreds of tasks generated by each annotator, or the construct validity of the tasks actually measuring the specific ToM dimensions are often buried in appendices or not reported at all. This issue is especially pronounced when datasets are generated through a mix of human annotators and generative AI. This highlights the need to standardize ToM benchmark reporting and documentation to ensure transparency, consistency, and validity in ToM benchmarks.

\subsection{Evaluation Limitation}
\textbf{ToM tasks rely on third-person, static, and synthetic scenarios, overlooking the practical use of ToM in dynamic, real-world social interactions.} ToM capability enables one to extract social cues embedded in the complex, dynamic, and multi-modal environment to attribute various mental states to self and others when facilitating social interactions. Yet most of the existing ToM tasks take social interactions out of its dynamic context, and consist solely of presenting static social scenarios or stories to examine the respondent's social understanding of the synthatic scenario from a third-person perspective~\cite{byom2013theory,quesque2020theory}.~\citeauthor{quesque2020theory}[2020] found that 17 out of the 23 classic ToM measures they reviewed only examine children's ToM from a passive observer perspective (i.e., third-person perspective). 

As~\citeauthor{ma2023towards} [2023] pointed out, this has also been the case in ToM benchmark literature in AI--- 12 out of the 21 papers in their survey positioned LLM as a passive observer, with only three papers positioning LLM as an active agent in the benchmark~\cite{ma2023towards}. Additionally,~\citeauthor{ma2023towards}[2023] highlighted the lack of benchmarks encompassing both the physical and social environments, overlooking other physical and spatial relationships between agents and the object as well as intrinsic motivations on the agent side~\cite{ma2023towards}. Ability to understand social scenario is not the exact equivalent or valid predictor of one's ability to engage in actual social interactions, especially when such scenarios are taken out of the social contexts. For instance, even in static social story tasks, \citeauthor{gu2024simpletom} [2024] found that while most LLMs could accurately infer characters’ mental states, they often failed to predict corresponding behaviors and performed worse when judging the reasonableness of those behaviors.~\citeauthor{riemer2025positiontheorymindbenchmarks} [2025] found that top-performing LLMs may possess strong literal ToM capability in predicting others' behaviors, they tend to struggle with functional ToM--- the ability to adapt to agents through rational responses based on valid predictions from literal ToM. This has spurred a paradigm shift to encourage the design of ToM tasks based on actual social interactions from a first- or second-person perspective in both psychology and AI literature~\cite{quesque2020theory,hou2024entering,zhou2023sotopia}. 

\section{Towards User-Centered Theory of Mind Benchmark for LLMs}
In the previous section, we summarized key limitations of appropriating human-intended ToM tasks to benchmark LLM's ToM capability based on existing literature: (1)~\textbf{Theoretical limitation}: ToM is a multi-faceted construct yet it has been mostly measured on one dimension, (2)~\textbf{Methodological limitation}: Many ToM tasks lack construct validity and present mixed or lack of reports of test validity and reliability, (3)~\textbf{Evaluation limitation}: ToM tasks rely on third-person, static, and synthetic scenarios, overlooking the practical use of ToM in dynamic, real-world social interactions. Our goal in surfacing and summarizing these limitations across psychology and AI literature is not to suggest abandoning ToM tasks entirely in LLM benchmarking but to highlight the challenges of repurposing them without further scrutiny in generating broad claims about LLM's general ToM capabilities. 

In the process of summarizing and highlighting these limitations, we noticed a recurring pattern across ToM benchmark work: the limited role assigned to humans in ToM benchmarks. Rather than participating as actual end-users of LLM-powered AI applications, humans primarily serve as annotators to generate ToM tasks or provide baseline measurements to benchmark LLM's ToM capabilities. From a user-centered perspective, we must ask: even if LLMs eventually match human ToM capabilities, these models will ultimately power user-facing applications, so shouldn't user perspectives, preferences, and needs inform ToM benchmark design? In the rest of this section, we take an HCI perspective to explore research opportunities and challenges for designing towards user-centered ToM benchmarks for LLMs. 


\subsection{Defining LLM ToM From a User-Centered Perspective}
Recognizing the theoretical limitations of existing ToM benchmarks, recent work has proposed more comprehensive evaluations that go beyond false-belief reasoning to cover multiple ToM dimensions identified in psychology~\cite{beaudoin2020systematic}, such as beliefs, desires, and emotions~\cite[e.g.,][]{chen2024tombench}. While such efforts represent important progress in understanding LLMs’ broader ToM capabilities, they still rest on a foundational assumption:~\textit{that frameworks developed for human social cognition are directly applicable to machines.} ToM benchmarks for LLMs--- and LLM benchmarks in general--- have often borrowed psychology theories, frameworks, constructs, and measurements to evaluate LLM's human-like capabilities. Though this provides a useful foundation, such adaptations also carry over the original goals and assumptions of those theories, which~\textit{may not align with the realities of AI design and deployment.} For instance, ToM tasks in developmental psychology were primarily designed to identify deficits in children's social reasoning in interpreting others’ mental states~\cite{baron1985does}—--not to define general-purpose models of social competence applicable to artificial agents interacting with humans in real life. 

LLMs are mostly used to power user-facing AI applications, so in practice, it may matter less whether LLMs possess ToM reasoning capabilities and more about the type of downstream behaviors enabled by LLM's ToM capabilities during human-AI interactions~\cite{gu2024simpletom}. In psychology, dimensions of ToM capabilities are good predictors for desirable human social behaviors, but this does not always translate to AI---human-AI interactions differ fundamentally from human-human interactions and thus users may have distinct expectations, preferences, and needs when it comes to AI's ToM-enabled behaviors. Not all ToM-enabled behaviors are desired and needed in every AI applications---~\citeauthor{borg2024required} [2024] pointed out that mot all empathic behaviors that fall under the umbrella capability of ``empathy'' will be needed and preferred for different empathic AI applications. Certain ToM-enabled behaviors that are considered socially adept in humans may instead elicit users' discomfort, distrust, or unease when exhibited by LLMs. For instance, an AI that predicts a user's intentions or thoughts too accurately might feel intrusive, raising concerns about user privacy. Similarly, AI systems that recognize and mimic a user’s emotions too well might come across as eerie or manipulative rather than empathetic. Some ToM-enabled behaviors may be unnecessary—--or even counterproductive—--in particular AI application contexts, such as productivity tools or navigation systems, where users prioritize efficiency and reliability over social attunement.

Taking a user-centered perspective, we urge researchers to rethink the definition of ToM in LLM benchmarks by moving beyond ``mimicking human behaviors'' through adapting psychology theories. Instead,~\textbf{we advocate for grounding the design of ToM benchmarks in empirical HCI studies that surface the kinds of ToM-enabled AI behaviors that users actually desire and need in real-world interactions.} This would require close collaboration between AI and HCI researchers to envision and implement ToM-enabled AI behaviors across diverse application contexts, conduct user studies or co-design sessions to understand people's interactions and experiences with such ToM-enabled AI systems, and translate those insights into measurable ToM dimensions for designing user-centered ToM benchmarks. Each portion presents its own unique challenges, the most difficult of which is the distillation of a comprehensive and easily-accessible ToM benchmark that meaningfully reflects the diverse user preferences and needs across various human-AI interaction contexts.

\subsection{Benchmarking LLM ToM in Dynamic and Interactional Social Contexts}
Several recent studies have proposed new approaches for aligning ToM benchmarks more closely with real world social contexts, in response to the limitations of evaluating LLMs using synthetic social scenarios. These approaches include leveraging natural human-human conversation dialogue to generate social scenarios in ToM benchmarks~\cite{soubki2024views,chan2024negotiationtom,kim2023fantom}, expanding the test modality to include both multiple-choice question-answering and free-form responses~\cite{kim2023fantom,chan2024negotiationtom}, as well as converting third-person perspective ToM tasks to first-person perspectives in ToM benchmarks~\cite{hou2024entering}. Open-ended interaction environment such as Sotopia~\cite{zhou2023sotopia} provides opportunities to assess LLM's ToM in more dynamic, socially complex and active interlocutor perspective. Additionally,~\citeauthor{shi2024muma}[2024] created multi-modal ToM benchmark that enables video and text descriptions of people's multi-modal behavior in realistic household environment to probe LLMs in answering about people's goals and beliefs. Work like~\citeauthor{yerukola2024pope} [2024] has also highlighted the importance to understand LLMs' capability in interpreting and responding to human intentions beyond the literal meaning of words to achieve ``functional ToM''~\cite{riemer2025positiontheorymindbenchmarks}.

As this effort towards more socially situated ToM benchmarks continues, we also want to reflect on the definition and criteria of LLMs ``passing'' ToM benchmarks when situated in more interactive and dynamic contexts. ToM has traditionally been viewed as a static construct that can be measured through the one-shot ``correctness'' of one's understanding of social cues through multiple-choice questions in ToM tasks. However, through the lens of Mutual Theory of Mind (MToM)~\cite{wang2022mutual,wang2024mutual,wang2021towards}, social interaction is iterative, sometimes requiring multiple back-and-forth between two parties through ToM construction, recognition, and revision for one to achieve the correct understanding of the other's mental states. As described by~\citeauthor{wang2022mutual} [2022] in their MToM framework, each turn of the communication can offer richer social signals through communication feedback, which builds upon the ToM inferences made from the previous turn to eventually arrive at the ``correct'' social understanding and attribution of mental states.

\textbf{In this light, when AI systems are embedded in dynamic, interactive environments, should we assess their ToM based on one-shot inference accuracy, or on their ability to iteratively refine their understanding in response to ongoing feedback?} If ToM is fundamentally about understanding and predicting others’ mental states in dynamic social environments, then~\textbf{a more meaningful benchmark should account for how well an AI system navigates the iterative nature of real-world social exchanges}. This shift in evaluation criteria would move beyond static correctness toward assessing an AI’s ToM based on its adaptability, responsiveness, and ability to integrate evolving social information, all of which are key components to human social intelligence. While readily quantifiable metrics--- such as the number of conversational turns required for accurate inference—-- offer a starting point, more nuanced measures that track improvements in inference quality based on individual user's feedback may provide deeper insight. Given the increasing deployment of LLM-powered AI applications in global contexts, such nuanced measures will need to include assessments on how well these systems can iteratively infer users' mental states when interacting with users across diverse demographic and cultural backgrounds--- something that even humans struggle with during cross-cultural communications. This adds another layer of complexity to ToM benchmark design, requiring methodological innovation that balances the depth of user-centered evaluation with the scalability required for robust ToM assessment.





\section{Conclusion}
In this position paper, we outlined and summarized limitations of the popular approach in appropriating ToM tasks designed to evaluate children's ToM to benchmark LLM's ToM. Drawing upon existing psychology and AI literature, we argue that these limitations already exist in the original human-intended ToM tasks, and hence persisted and exacerbated when appropriated as LLM benchmarks. Specifically, we summarized three key limitations: (1) Theoretical limitation: ToM is a multi-faceted construct yet it has been mostly measured on one dimension, (2) Methodological limitation: Many ToM tasks lack construct validity and present mixed or lack of reports of test validity and reliability, (3) Evaluation limitation: ToM tasks rely on third-person, static, and synthetic scenarios, overlooking the practical use of ToM in dynamic, real-world social interactions. By identifying these limitations, we caution AI researchers against blind adoption of these ToM tasks and to draw claims about LLM's general ToM capability based on LLM passing such ToM tasks. Based on these limitations, we proposed the future direction towards designing user-centered ToM benchmark for LLMs. We discuss potential opportunities and challenges in this direction and encourage researchers to rethink the definition of LLM ToM based on user needs and preferences, as well as reflecting on the criteria of LLM benchmark in dynamic and interactive social contexts. 







\bibliographystyle{ACM-Reference-Format}
\bibliography{reference}


\begin{thebibliography}{49}


\ifx \showCODEN    \undefined \def \showCODEN     #1{\unskip}     \fi
\ifx \showDOI      \undefined \def \showDOI       #1{#1}\fi
\ifx \showISBNx    \undefined \def \showISBNx     #1{\unskip}     \fi
\ifx \showISBNxiii \undefined \def \showISBNxiii  #1{\unskip}     \fi
\ifx \showISSN     \undefined \def \showISSN      #1{\unskip}     \fi
\ifx \showLCCN     \undefined \def \showLCCN      #1{\unskip}     \fi
\ifx \shownote     \undefined \def \shownote      #1{#1}          \fi
\ifx \showarticletitle \undefined \def \showarticletitle #1{#1}   \fi
\ifx \showURL      \undefined \def \showURL       {\relax}        \fi
\providecommand\bibfield[2]{#2}
\providecommand\bibinfo[2]{#2}
\providecommand\natexlab[1]{#1}
\providecommand\showeprint[2][]{arXiv:#2}

\bibitem[Ahmadi et~al\mbox{.}(2015)]%
        {ahmadi2015validity}
\bibfield{author}{\bibinfo{person}{Seyyede Zohreh~Ziatabar Ahmadi}, \bibinfo{person}{Shohreh Jalaie}, {and} \bibinfo{person}{Hassan Ashayeri}.} \bibinfo{year}{2015}\natexlab{}.
\newblock \showarticletitle{Validity and reliability of published comprehensive theory of mind tests for normal preschool children: A systematic review}.
\newblock \bibinfo{journal}{\emph{Iranian journal of psychiatry}} \bibinfo{volume}{10}, \bibinfo{number}{4} (\bibinfo{year}{2015}), \bibinfo{pages}{214}.
\newblock


\bibitem[Baron-Cohen(1999)]%
        {baron1999evolution}
\bibfield{author}{\bibinfo{person}{Simon Baron-Cohen}.} \bibinfo{year}{1999}\natexlab{}.
\newblock \bibinfo{booktitle}{\emph{The evolution of a theory of mind}}.
\newblock \bibinfo{publisher}{na}.
\newblock


\bibitem[Baron-Cohen(2000)]%
        {baron2000theory}
\bibfield{author}{\bibinfo{person}{Simon Baron-Cohen}.} \bibinfo{year}{2000}\natexlab{}.
\newblock \showarticletitle{Theory of mind and autism: A review}.
\newblock \bibinfo{journal}{\emph{International review of research in mental retardation}}  \bibinfo{volume}{23} (\bibinfo{year}{2000}), \bibinfo{pages}{169--184}.
\newblock


\bibitem[Baron-Cohen et~al\mbox{.}(1985)]%
        {baron1985does}
\bibfield{author}{\bibinfo{person}{Simon Baron-Cohen}, \bibinfo{person}{Alan~M Leslie}, {and} \bibinfo{person}{Uta Frith}.} \bibinfo{year}{1985}\natexlab{}.
\newblock \showarticletitle{Does the autistic child have a “theory of mind”?}
\newblock \bibinfo{journal}{\emph{Cognition}} \bibinfo{volume}{21}, \bibinfo{number}{1} (\bibinfo{year}{1985}), \bibinfo{pages}{37--46}.
\newblock


\bibitem[Baron-Cohen et~al\mbox{.}(1999)]%
        {baron1999recognition}
\bibfield{author}{\bibinfo{person}{Simon Baron-Cohen}, \bibinfo{person}{Michelle O'riordan}, \bibinfo{person}{Valerie Stone}, \bibinfo{person}{Rosie Jones}, {and} \bibinfo{person}{Kate Plaisted}.} \bibinfo{year}{1999}\natexlab{}.
\newblock \showarticletitle{Recognition of faux pas by normally developing children and children with Asperger syndrome or high-functioning autism}.
\newblock \bibinfo{journal}{\emph{Journal of autism and developmental disorders}}  \bibinfo{volume}{29} (\bibinfo{year}{1999}), \bibinfo{pages}{407--418}.
\newblock


\bibitem[Baron-Cohen et~al\mbox{.}(2001)]%
        {baron2001reading}
\bibfield{author}{\bibinfo{person}{Simon Baron-Cohen}, \bibinfo{person}{Sally Wheelwright}, \bibinfo{person}{Jacqueline Hill}, \bibinfo{person}{Yogini Raste}, {and} \bibinfo{person}{Ian Plumb}.} \bibinfo{year}{2001}\natexlab{}.
\newblock \showarticletitle{The “Reading the Mind in the Eyes” test revised version: A study with normal adults, and adults with Asperger syndrome or high-functioning autism}.
\newblock \bibinfo{journal}{\emph{Journal of child psychology and psychiatry}} \bibinfo{volume}{42}, \bibinfo{number}{2} (\bibinfo{year}{2001}), \bibinfo{pages}{241--251}.
\newblock


\bibitem[Beaudoin et~al\mbox{.}(2020)]%
        {beaudoin2020systematic}
\bibfield{author}{\bibinfo{person}{Cindy Beaudoin}, \bibinfo{person}{{\'E}lizabel Leblanc}, \bibinfo{person}{Charlotte Gagner}, {and} \bibinfo{person}{Miriam~H Beauchamp}.} \bibinfo{year}{2020}\natexlab{}.
\newblock \showarticletitle{Systematic review and inventory of theory of mind measures for young children}.
\newblock \bibinfo{journal}{\emph{Frontiers in psychology}}  \bibinfo{volume}{10} (\bibinfo{year}{2020}), \bibinfo{pages}{2905}.
\newblock


\bibitem[Borg and Read(2024)]%
        {borg2024required}
\bibfield{author}{\bibinfo{person}{Jana~Schaich Borg} {and} \bibinfo{person}{Hannah Read}.} \bibinfo{year}{2024}\natexlab{}.
\newblock \showarticletitle{What Is Required for Empathic AI? It Depends, and Why That Matters for AI Developers and Users}. In \bibinfo{booktitle}{\emph{Proceedings of the AAAI/ACM Conference on AI, Ethics, and Society}}, Vol.~\bibinfo{volume}{7}. \bibinfo{pages}{1306--1318}.
\newblock


\bibitem[Bubeck et~al\mbox{.}(2023)]%
        {bubeck2023sparks}
\bibfield{author}{\bibinfo{person}{S{\'e}bastien Bubeck}, \bibinfo{person}{Varun Chandrasekaran}, \bibinfo{person}{Ronen Eldan}, \bibinfo{person}{Johannes Gehrke}, \bibinfo{person}{Eric Horvitz}, \bibinfo{person}{Ece Kamar}, \bibinfo{person}{Peter Lee}, \bibinfo{person}{Yin~Tat Lee}, \bibinfo{person}{Yuanzhi Li}, \bibinfo{person}{Scott Lundberg}, {et~al\mbox{.}}} \bibinfo{year}{2023}\natexlab{}.
\newblock \showarticletitle{Sparks of artificial general intelligence: Early experiments with gpt-4}.
\newblock \bibinfo{journal}{\emph{arXiv preprint arXiv:2303.12712}} (\bibinfo{year}{2023}).
\newblock


\bibitem[Byom and Mutlu(2013)]%
        {byom2013theory}
\bibfield{author}{\bibinfo{person}{Lindsey~J Byom} {and} \bibinfo{person}{Bilge Mutlu}.} \bibinfo{year}{2013}\natexlab{}.
\newblock \showarticletitle{Theory of mind: Mechanisms, methods, and new directions}.
\newblock \bibinfo{journal}{\emph{Frontiers in human neuroscience}}  \bibinfo{volume}{7} (\bibinfo{year}{2013}), \bibinfo{pages}{413}.
\newblock


\bibitem[Chan et~al\mbox{.}(2024)]%
        {chan2024negotiationtom}
\bibfield{author}{\bibinfo{person}{Chunkit Chan}, \bibinfo{person}{Cheng Jiayang}, \bibinfo{person}{Yauwai Yim}, \bibinfo{person}{Zheye Deng}, \bibinfo{person}{Wei Fan}, \bibinfo{person}{Haoran Li}, \bibinfo{person}{Xin Liu}, \bibinfo{person}{Hongming Zhang}, \bibinfo{person}{Weiqi Wang}, {and} \bibinfo{person}{Yangqiu Song}.} \bibinfo{year}{2024}\natexlab{}.
\newblock \showarticletitle{NegotiationToM: A Benchmark for Stress-testing Machine Theory of Mind on Negotiation Surrounding}.
\newblock \bibinfo{journal}{\emph{arXiv preprint arXiv:2404.13627}} (\bibinfo{year}{2024}).
\newblock


\bibitem[Chen et~al\mbox{.}(2024)]%
        {chen2024tombench}
\bibfield{author}{\bibinfo{person}{Zhuang Chen}, \bibinfo{person}{Jincenzi Wu}, \bibinfo{person}{Jinfeng Zhou}, \bibinfo{person}{Bosi Wen}, \bibinfo{person}{Guanqun Bi}, \bibinfo{person}{Gongyao Jiang}, \bibinfo{person}{Yaru Cao}, \bibinfo{person}{Mengting Hu}, \bibinfo{person}{Yunghwei Lai}, \bibinfo{person}{Zexuan Xiong}, {et~al\mbox{.}}} \bibinfo{year}{2024}\natexlab{}.
\newblock \showarticletitle{ToMBench: Benchmarking Theory of Mind in Large Language Models}.
\newblock \bibinfo{journal}{\emph{arXiv preprint arXiv:2402.15052}} (\bibinfo{year}{2024}).
\newblock


\bibitem[Chiang et~al\mbox{.}(2024)]%
        {chiang2024chatbot}
\bibfield{author}{\bibinfo{person}{Wei-Lin Chiang}, \bibinfo{person}{Lianmin Zheng}, \bibinfo{person}{Ying Sheng}, \bibinfo{person}{Anastasios~Nikolas Angelopoulos}, \bibinfo{person}{Tianle Li}, \bibinfo{person}{Dacheng Li}, \bibinfo{person}{Hao Zhang}, \bibinfo{person}{Banghua Zhu}, \bibinfo{person}{Michael Jordan}, \bibinfo{person}{Joseph~E Gonzalez}, {et~al\mbox{.}}} \bibinfo{year}{2024}\natexlab{}.
\newblock \showarticletitle{Chatbot arena: An open platform for evaluating llms by human preference}.
\newblock \bibinfo{journal}{\emph{arXiv preprint arXiv:2403.04132}} (\bibinfo{year}{2024}).
\newblock


\bibitem[Fu et~al\mbox{.}(2023)]%
        {fu2023systematic}
\bibfield{author}{\bibinfo{person}{I-Ning Fu}, \bibinfo{person}{Kuan-Lin Chen}, \bibinfo{person}{Meng-Ru Liu}, \bibinfo{person}{Dai-Rong Jiang}, \bibinfo{person}{Ching-Lin Hsieh}, {and} \bibinfo{person}{Shih-Chieh Lee}.} \bibinfo{year}{2023}\natexlab{}.
\newblock \showarticletitle{A systematic review of measures of theory of mind for children}.
\newblock \bibinfo{journal}{\emph{Developmental Review}}  \bibinfo{volume}{67} (\bibinfo{year}{2023}), \bibinfo{pages}{101061}.
\newblock


\bibitem[Gu et~al\mbox{.}(2024)]%
        {gu2024simpletom}
\bibfield{author}{\bibinfo{person}{Yuling Gu}, \bibinfo{person}{Oyvind Tafjord}, \bibinfo{person}{Hyunwoo Kim}, \bibinfo{person}{Jared Moore}, \bibinfo{person}{Ronan~Le Bras}, \bibinfo{person}{Peter Clark}, {and} \bibinfo{person}{Yejin Choi}.} \bibinfo{year}{2024}\natexlab{}.
\newblock \showarticletitle{SimpleToM: Exposing the Gap between Explicit ToM Inference and Implicit ToM Application in LLMs}.
\newblock \bibinfo{journal}{\emph{arXiv preprint arXiv:2410.13648}} (\bibinfo{year}{2024}).
\newblock


\bibitem[Happ{\'e}(1993)]%
        {happe1993communicative}
\bibfield{author}{\bibinfo{person}{Francesca~GE Happ{\'e}}.} \bibinfo{year}{1993}\natexlab{}.
\newblock \showarticletitle{Communicative competence and theory of mind in autism: A test of relevance theory}.
\newblock \bibinfo{journal}{\emph{Cognition}} \bibinfo{volume}{48}, \bibinfo{number}{2} (\bibinfo{year}{1993}), \bibinfo{pages}{101--119}.
\newblock


\bibitem[Hayward and Homer(2017)]%
        {hayward2017reliability}
\bibfield{author}{\bibinfo{person}{Elizabeth~O Hayward} {and} \bibinfo{person}{Bruce~D Homer}.} \bibinfo{year}{2017}\natexlab{}.
\newblock \showarticletitle{Reliability and validity of advanced theory-of-mind measures in middle childhood and adolescence}.
\newblock \bibinfo{journal}{\emph{British Journal of Developmental Psychology}} \bibinfo{volume}{35}, \bibinfo{number}{3} (\bibinfo{year}{2017}), \bibinfo{pages}{454--462}.
\newblock


\bibitem[Hou et~al\mbox{.}(2024)]%
        {hou2024entering}
\bibfield{author}{\bibinfo{person}{Guiyang Hou}, \bibinfo{person}{Wenqi Zhang}, \bibinfo{person}{Yongliang Shen}, \bibinfo{person}{Zeqi Tan}, \bibinfo{person}{Sihao Shen}, {and} \bibinfo{person}{Weiming Lu}.} \bibinfo{year}{2024}\natexlab{}.
\newblock \showarticletitle{Entering Real Social World! Benchmarking the Theory of Mind and Socialization Capabilities of LLMs from a First-person Perspective}.
\newblock \bibinfo{journal}{\emph{arXiv preprint arXiv:2410.06195}} (\bibinfo{year}{2024}).
\newblock


\bibitem[Huang et~al\mbox{.}(2025)]%
        {huang2025apathetic}
\bibfield{author}{\bibinfo{person}{Jen-tse Huang}, \bibinfo{person}{Man~Ho Lam}, \bibinfo{person}{Eric~John Li}, \bibinfo{person}{Shujie Ren}, \bibinfo{person}{Wenxuan Wang}, \bibinfo{person}{Wenxiang Jiao}, \bibinfo{person}{Zhaopeng Tu}, {and} \bibinfo{person}{Michael~R Lyu}.} \bibinfo{year}{2025}\natexlab{}.
\newblock \showarticletitle{Apathetic or Empathetic? Evaluating LLMs' Emotional Alignments with Humans}.
\newblock \bibinfo{journal}{\emph{Advances in Neural Information Processing Systems}}  \bibinfo{volume}{37} (\bibinfo{year}{2025}), \bibinfo{pages}{97053--97087}.
\newblock


\bibitem[Kim et~al\mbox{.}(2023)]%
        {kim2023fantom}
\bibfield{author}{\bibinfo{person}{Hyunwoo Kim}, \bibinfo{person}{Melanie Sclar}, \bibinfo{person}{Xuhui Zhou}, \bibinfo{person}{Ronan~Le Bras}, \bibinfo{person}{Gunhee Kim}, \bibinfo{person}{Yejin Choi}, {and} \bibinfo{person}{Maarten Sap}.} \bibinfo{year}{2023}\natexlab{}.
\newblock \showarticletitle{FANToM: A benchmark for stress-testing machine theory of mind in interactions}.
\newblock \bibinfo{journal}{\emph{arXiv preprint arXiv:2310.15421}} (\bibinfo{year}{2023}).
\newblock


\bibitem[Kosinski({[n.\,d.]})]%
        {kosinski2023theory}
\bibfield{author}{\bibinfo{person}{Michal Kosinski}.} \bibinfo{year}{[n.\,d.]}\natexlab{}.
\newblock \showarticletitle{Theory of mind may have spontaneously emerged in large language models}.
\newblock  (\bibinfo{year}{[n.\,d.]}).
\newblock


\bibitem[Le et~al\mbox{.}(2019)]%
        {le2019revisiting}
\bibfield{author}{\bibinfo{person}{Matthew Le}, \bibinfo{person}{Y-Lan Boureau}, {and} \bibinfo{person}{Maximilian Nickel}.} \bibinfo{year}{2019}\natexlab{}.
\newblock \showarticletitle{Revisiting the evaluation of theory of mind through question answering}. In \bibinfo{booktitle}{\emph{Proceedings of the 2019 Conference on Empirical Methods in Natural Language Processing and the 9th International Joint Conference on Natural Language Processing (EMNLP-IJCNLP)}}. \bibinfo{pages}{5872--5877}.
\newblock


\bibitem[Ma et~al\mbox{.}(2023)]%
        {ma2023towards}
\bibfield{author}{\bibinfo{person}{Ziqiao Ma}, \bibinfo{person}{Jacob Sansom}, \bibinfo{person}{Run Peng}, {and} \bibinfo{person}{Joyce Chai}.} \bibinfo{year}{2023}\natexlab{}.
\newblock \showarticletitle{Towards a holistic landscape of situated theory of mind in large language models}.
\newblock \bibinfo{journal}{\emph{arXiv preprint arXiv:2310.19619}} (\bibinfo{year}{2023}).
\newblock


\bibitem[Milton(2012)]%
        {milton2012ontological}
\bibfield{author}{\bibinfo{person}{Damian~EM Milton}.} \bibinfo{year}{2012}\natexlab{}.
\newblock \showarticletitle{On the ontological status of autism: The ‘double empathy problem’}.
\newblock \bibinfo{journal}{\emph{Disability \& society}} \bibinfo{volume}{27}, \bibinfo{number}{6} (\bibinfo{year}{2012}), \bibinfo{pages}{883--887}.
\newblock


\bibitem[Oakley et~al\mbox{.}(2016)]%
        {oakley2016theory}
\bibfield{author}{\bibinfo{person}{Beth~FM Oakley}, \bibinfo{person}{Rebecca Brewer}, \bibinfo{person}{Geoffrey Bird}, {and} \bibinfo{person}{Caroline Catmur}.} \bibinfo{year}{2016}\natexlab{}.
\newblock \showarticletitle{Theory of mind is not theory of emotion: A cautionary note on the Reading the Mind in the Eyes Test.}
\newblock \bibinfo{journal}{\emph{Journal of abnormal psychology}} \bibinfo{volume}{125}, \bibinfo{number}{6} (\bibinfo{year}{2016}), \bibinfo{pages}{818}.
\newblock


\bibitem[Premack and Woodruff(1978)]%
        {premack1978does}
\bibfield{author}{\bibinfo{person}{David Premack} {and} \bibinfo{person}{Guy Woodruff}.} \bibinfo{year}{1978}\natexlab{}.
\newblock \showarticletitle{Does the chimpanzee have a theory of mind?}
\newblock \bibinfo{journal}{\emph{Behavioral and brain sciences}} \bibinfo{volume}{1}, \bibinfo{number}{4} (\bibinfo{year}{1978}), \bibinfo{pages}{515--526}.
\newblock


\bibitem[Quesque and Rossetti(2020)]%
        {quesque2020theory}
\bibfield{author}{\bibinfo{person}{Fran{\c{c}}ois Quesque} {and} \bibinfo{person}{Yves Rossetti}.} \bibinfo{year}{2020}\natexlab{}.
\newblock \showarticletitle{What do theory-of-mind tasks actually measure? Theory and practice}.
\newblock \bibinfo{journal}{\emph{Perspectives on Psychological Science}} \bibinfo{volume}{15}, \bibinfo{number}{2} (\bibinfo{year}{2020}), \bibinfo{pages}{384--396}.
\newblock


\bibitem[Rakoczy(2022)]%
        {rakoczy2022foundations}
\bibfield{author}{\bibinfo{person}{Hannes Rakoczy}.} \bibinfo{year}{2022}\natexlab{}.
\newblock \showarticletitle{Foundations of theory of mind and its development in early childhood}.
\newblock \bibinfo{journal}{\emph{Nature Reviews Psychology}} \bibinfo{volume}{1}, \bibinfo{number}{4} (\bibinfo{year}{2022}), \bibinfo{pages}{223--235}.
\newblock


\bibitem[Riemer et~al\mbox{.}(2025)]%
        {riemer2025positiontheorymindbenchmarks}
\bibfield{author}{\bibinfo{person}{Matthew Riemer}, \bibinfo{person}{Zahra Ashktorab}, \bibinfo{person}{Djallel Bouneffouf}, \bibinfo{person}{Payel Das}, \bibinfo{person}{Miao Liu}, \bibinfo{person}{Justin~D. Weisz}, {and} \bibinfo{person}{Murray Campbell}.} \bibinfo{year}{2025}\natexlab{}.
\newblock \bibinfo{title}{Position: Theory of Mind Benchmarks are Broken for Large Language Models}.
\newblock
\newblock
\showeprint[arxiv]{2412.19726}~[cs.AI]
\urldef\tempurl%
\url{https://arxiv.org/abs/2412.19726}
\showURL{%
\tempurl}


\bibitem[Sap et~al\mbox{.}(2022)]%
        {sap2022neural}
\bibfield{author}{\bibinfo{person}{Maarten Sap}, \bibinfo{person}{Ronan LeBras}, \bibinfo{person}{Daniel Fried}, {and} \bibinfo{person}{Yejin Choi}.} \bibinfo{year}{2022}\natexlab{}.
\newblock \showarticletitle{Neural theory-of-mind? on the limits of social intelligence in large lms}.
\newblock \bibinfo{journal}{\emph{arXiv preprint arXiv:2210.13312}} (\bibinfo{year}{2022}).
\newblock


\bibitem[Sap et~al\mbox{.}(2019)]%
        {sap2019socialiqa}
\bibfield{author}{\bibinfo{person}{Maarten Sap}, \bibinfo{person}{Hannah Rashkin}, \bibinfo{person}{Derek Chen}, \bibinfo{person}{Ronan LeBras}, {and} \bibinfo{person}{Yejin Choi}.} \bibinfo{year}{2019}\natexlab{}.
\newblock \showarticletitle{Socialiqa: Commonsense reasoning about social interactions}.
\newblock \bibinfo{journal}{\emph{arXiv preprint arXiv:1904.09728}} (\bibinfo{year}{2019}).
\newblock


\bibitem[Scassellati(2001)]%
        {scassellati2001foundations}
\bibfield{author}{\bibinfo{person}{Brian~M Scassellati}.} \bibinfo{year}{2001}\natexlab{}.
\newblock \emph{\bibinfo{title}{Foundations for a Theory of Mind for a Humanoid Robot}}.
\newblock \bibinfo{thesistype}{Ph.\,D. Dissertation}. \bibinfo{school}{Massachusetts Institute of Technology}.
\newblock


\bibitem[Shapira et~al\mbox{.}(2023a)]%
        {shapira2023clever}
\bibfield{author}{\bibinfo{person}{Natalie Shapira}, \bibinfo{person}{Mosh Levy}, \bibinfo{person}{Seyed~Hossein Alavi}, \bibinfo{person}{Xuhui Zhou}, \bibinfo{person}{Yejin Choi}, \bibinfo{person}{Yoav Goldberg}, \bibinfo{person}{Maarten Sap}, {and} \bibinfo{person}{Vered Shwartz}.} \bibinfo{year}{2023}\natexlab{a}.
\newblock \showarticletitle{Clever hans or neural theory of mind? stress testing social reasoning in large language models}.
\newblock \bibinfo{journal}{\emph{arXiv preprint arXiv:2305.14763}} (\bibinfo{year}{2023}).
\newblock


\bibitem[Shapira et~al\mbox{.}(2023b)]%
        {shapira2023well}
\bibfield{author}{\bibinfo{person}{Natalie Shapira}, \bibinfo{person}{Guy Zwirn}, {and} \bibinfo{person}{Yoav Goldberg}.} \bibinfo{year}{2023}\natexlab{b}.
\newblock \showarticletitle{How well do large language models perform on faux pas tests?}. In \bibinfo{booktitle}{\emph{Findings of the Association for Computational Linguistics: ACL 2023}}. \bibinfo{pages}{10438--10451}.
\newblock


\bibitem[Shi et~al\mbox{.}(2024)]%
        {shi2024muma}
\bibfield{author}{\bibinfo{person}{Haojun Shi}, \bibinfo{person}{Suyu Ye}, \bibinfo{person}{Xinyu Fang}, \bibinfo{person}{Chuanyang Jin}, \bibinfo{person}{Leyla Isik}, \bibinfo{person}{Yen-Ling Kuo}, {and} \bibinfo{person}{Tianmin Shu}.} \bibinfo{year}{2024}\natexlab{}.
\newblock \showarticletitle{Muma-tom: Multi-modal multi-agent theory of mind}.
\newblock \bibinfo{journal}{\emph{arXiv preprint arXiv:2408.12574}} (\bibinfo{year}{2024}).
\newblock


\bibitem[Sorin et~al\mbox{.}(2024)]%
        {sorin2024large}
\bibfield{author}{\bibinfo{person}{Vera Sorin}, \bibinfo{person}{Dana Brin}, \bibinfo{person}{Yiftach Barash}, \bibinfo{person}{Eli Konen}, \bibinfo{person}{Alexander Charney}, \bibinfo{person}{Girish Nadkarni}, {and} \bibinfo{person}{Eyal Klang}.} \bibinfo{year}{2024}\natexlab{}.
\newblock \showarticletitle{Large Language Models and Empathy: Systematic Review}.
\newblock \bibinfo{journal}{\emph{Journal of Medical Internet Research}}  \bibinfo{volume}{26} (\bibinfo{year}{2024}), \bibinfo{pages}{e52597}.
\newblock


\bibitem[Soubki et~al\mbox{.}(2024)]%
        {soubki2024views}
\bibfield{author}{\bibinfo{person}{Adil Soubki}, \bibinfo{person}{John Murzaku}, \bibinfo{person}{Arash~Yousefi Jordehi}, \bibinfo{person}{Peter Zeng}, \bibinfo{person}{Magdalena Markowska}, \bibinfo{person}{Seyed~Abolghasem Mirroshandel}, {and} \bibinfo{person}{Owen Rambow}.} \bibinfo{year}{2024}\natexlab{}.
\newblock \showarticletitle{Views Are My Own, But Also Yours: Benchmarking Theory of Mind using Common Ground}.
\newblock \bibinfo{journal}{\emph{arXiv preprint arXiv:2403.02451}} (\bibinfo{year}{2024}).
\newblock


\bibitem[Ullman(2023)]%
        {ullman2023large}
\bibfield{author}{\bibinfo{person}{Tomer Ullman}.} \bibinfo{year}{2023}\natexlab{}.
\newblock \showarticletitle{Large language models fail on trivial alterations to theory-of-mind tasks}.
\newblock \bibinfo{journal}{\emph{arXiv preprint arXiv:2302.08399}} (\bibinfo{year}{2023}).
\newblock


\bibitem[Wang(2024)]%
        {wang2024mutual}
\bibfield{author}{\bibinfo{person}{Qiaosi Wang}.} \bibinfo{year}{2024}\natexlab{}.
\newblock \emph{\bibinfo{title}{MUTUAL THEORY OF MIND FOR HUMAN-AI COMMUNICATION IN AI-MEDIATED SOCIAL INTERACTION}}.
\newblock \bibinfo{thesistype}{Ph.\,D. Dissertation}. \bibinfo{school}{Georgia Institute of Technology}.
\newblock


\bibitem[Wang and Goel(2022)]%
        {wang2022mutual}
\bibfield{author}{\bibinfo{person}{Qiaosi Wang} {and} \bibinfo{person}{Ashok~K Goel}.} \bibinfo{year}{2022}\natexlab{}.
\newblock \showarticletitle{Mutual theory of mind for human-AI communication}.
\newblock \bibinfo{journal}{\emph{arXiv preprint arXiv:2210.03842}} (\bibinfo{year}{2022}).
\newblock


\bibitem[Wang et~al\mbox{.}(2021)]%
        {wang2021towards}
\bibfield{author}{\bibinfo{person}{Qiaosi Wang}, \bibinfo{person}{Koustuv Saha}, \bibinfo{person}{Eric Gregori}, \bibinfo{person}{David Joyner}, {and} \bibinfo{person}{Ashok Goel}.} \bibinfo{year}{2021}\natexlab{}.
\newblock \showarticletitle{Towards mutual theory of mind in human-ai interaction: How language reflects what students perceive about a virtual teaching assistant}. In \bibinfo{booktitle}{\emph{Proceedings of the 2021 CHI conference on human factors in computing systems}}. \bibinfo{pages}{1--14}.
\newblock


\bibitem[Welivita and Pu(2024)]%
        {welivita2024large}
\bibfield{author}{\bibinfo{person}{Anuradha Welivita} {and} \bibinfo{person}{Pearl Pu}.} \bibinfo{year}{2024}\natexlab{}.
\newblock \showarticletitle{Are Large Language Models More Empathetic than Humans?}
\newblock \bibinfo{journal}{\emph{arXiv preprint arXiv:2406.05063}} (\bibinfo{year}{2024}).
\newblock


\bibitem[Wellman(2018)]%
        {wellman2018theory}
\bibfield{author}{\bibinfo{person}{Henry~M Wellman}.} \bibinfo{year}{2018}\natexlab{}.
\newblock \showarticletitle{Theory of mind: The state of the art}.
\newblock \bibinfo{journal}{\emph{European Journal of Developmental Psychology}} \bibinfo{volume}{15}, \bibinfo{number}{6} (\bibinfo{year}{2018}), \bibinfo{pages}{728--755}.
\newblock


\bibitem[Xu et~al\mbox{.}(2024)]%
        {xu2024opentom}
\bibfield{author}{\bibinfo{person}{Hainiu Xu}, \bibinfo{person}{Runcong Zhao}, \bibinfo{person}{Lixing Zhu}, \bibinfo{person}{Jinhua Du}, {and} \bibinfo{person}{Yulan He}.} \bibinfo{year}{2024}\natexlab{}.
\newblock \showarticletitle{OpenToM: A Comprehensive Benchmark for Evaluating Theory-of-Mind Reasoning Capabilities of Large Language Models}.
\newblock \bibinfo{journal}{\emph{arXiv preprint arXiv:2402.06044}} (\bibinfo{year}{2024}).
\newblock


\bibitem[Yerukola et~al\mbox{.}(2024)]%
        {yerukola2024pope}
\bibfield{author}{\bibinfo{person}{Akhila Yerukola}, \bibinfo{person}{Saujas Vaduguru}, \bibinfo{person}{Daniel Fried}, {and} \bibinfo{person}{Maarten Sap}.} \bibinfo{year}{2024}\natexlab{}.
\newblock \showarticletitle{Is the pope catholic? yes, the pope is catholic. generative evaluation of non-literal intent resolution in LLMs}. In \bibinfo{booktitle}{\emph{Proceedings of the 62nd Annual Meeting of the Association for Computational Linguistics (Volume 2: Short Papers)}}. \bibinfo{pages}{265--275}.
\newblock


\bibitem[Ying et~al\mbox{.}(2025)]%
        {ying2025benchmarking}
\bibfield{author}{\bibinfo{person}{Lance Ying}, \bibinfo{person}{Katherine~M Collins}, \bibinfo{person}{Lionel Wong}, \bibinfo{person}{Ilia Sucholutsky}, \bibinfo{person}{Ryan Liu}, \bibinfo{person}{Adrian Weller}, \bibinfo{person}{Tianmin Shu}, \bibinfo{person}{Thomas~L Griffiths}, {and} \bibinfo{person}{Joshua~B Tenenbaum}.} \bibinfo{year}{2025}\natexlab{}.
\newblock \showarticletitle{On Benchmarking Human-Like Intelligence in Machines}.
\newblock \bibinfo{journal}{\emph{arXiv preprint arXiv:2502.20502}} (\bibinfo{year}{2025}).
\newblock


\bibitem[Zhang et~al\mbox{.}({[n.\,d.]})]%
        {zhangdeveloping}
\bibfield{author}{\bibinfo{person}{Haoran Zhang}, \bibinfo{person}{Ling Wang}, \bibinfo{person}{Zhihao Chen}, \bibinfo{person}{Yue Liu}, {and} \bibinfo{person}{Xinyi Li1~Jing Wu}.} \bibinfo{year}{[n.\,d.]}\natexlab{}.
\newblock \showarticletitle{Developing a Comprehensive Empathy Evaluation Benchmark for AI Systems}.
\newblock  (\bibinfo{year}{[n.\,d.]}).
\newblock


\bibitem[Zhou et~al\mbox{.}(2024)]%
        {zhou2023sotopia}
\bibfield{author}{\bibinfo{person}{Xuhui Zhou}, \bibinfo{person}{Hao Zhu}, \bibinfo{person}{Leena Mathur}, \bibinfo{person}{Ruohong Zhang}, \bibinfo{person}{Haofei Yu}, \bibinfo{person}{Zhengyang Qi}, \bibinfo{person}{Louis-Philippe Morency}, \bibinfo{person}{Yonatan Bisk}, \bibinfo{person}{Daniel Fried}, \bibinfo{person}{Graham Neubig}, {et~al\mbox{.}}} \bibinfo{year}{2024}\natexlab{}.
\newblock \showarticletitle{Sotopia: Interactive evaluation for social intelligence in language agents}. In \bibinfo{booktitle}{\emph{ICLR}}.
\newblock
\urldef\tempurl%
\url{https://arxiv.org/abs/2310.11667}
\showURL{%
\tempurl}


\bibitem[Zhu et~al\mbox{.}(2024)]%
        {zhu2024reading}
\bibfield{author}{\bibinfo{person}{Qihao Zhu}, \bibinfo{person}{Leah Chong}, \bibinfo{person}{Maria Yang}, {and} \bibinfo{person}{Jianxi Luo}.} \bibinfo{year}{2024}\natexlab{}.
\newblock \showarticletitle{Reading users’ minds from what they say: An investigation into llm-based empathic mental inference}. In \bibinfo{booktitle}{\emph{International Design Engineering Technical Conferences and Computers and Information in Engineering Conference}}, Vol.~\bibinfo{volume}{88407}. American Society of Mechanical Engineers, \bibinfo{pages}{V006T06A018}.
\newblock


\end{thebibliography}

\end{document}
\endinput